\documentclass{emulateapj}
\usepackage{color}
\usepackage{lineno}
\usepackage{amsmath}
\shorttitle{Broadband Early Afterglow of MAGIC-Burst GRB 190114C}
\shortauthors{Asano, Murase \& Toma}

\begin{document}
\pagenumbering{arabic}
\title{
Probing Particle Acceleration through Broadband Early Afterglow Emission\\ 
of MAGIC Gamma-Ray Burst GRB 190114C
}
\author{Katsuaki Asano\altaffilmark{1},
Kohta Murase\altaffilmark{2,3,4,5}, and Kenji Toma\altaffilmark{6,7}}
\email{asanok@icrr.u-tokyo.ac.jp}

\affil{\altaffilmark{1}Institute for Cosmic Ray Research, The University of Tokyo,
5-1-5 Kashiwanoha, Kashiwa, Chiba 277-8582, Japan;
asanok@icrr.u-tokyo.ac.jp}
\affil{\altaffilmark{2}Department of Physics, The Pennsylvania State University, University Park, Pennsylvania 16802, USA}
\affil{\altaffilmark{3}Department of Astronomy \& Astrophysics, The Pennsylvania State University, University Park, Pennsylvania 16802, USA}
\affil{\altaffilmark{4}Center for Multimessenger Astrophysics, Institute for Gravitation and the Cosmos, The Pennsylvania State University, University Park, Pennsylvania 16802, USA}
\affil{\altaffilmark{5}Center for Gravitational Physics, Yukawa Institute for Theoretical Physics, Kyoto University, Kyoto 606-8502, Japan}
\affil{\altaffilmark{6}Frontier Research Institute for Interdisciplinary Sciences, Tohoku University, Sendai 980-8578, Japan}
\affil{\altaffilmark{7}Astronomical Institute, Graduate School of Science, Tohoku University, Sendai 980-8578, Japan}
\date{Submitted; accepted}

\begin{abstract}
Major Atmospheric Gamma Imaging Cherenkov Telescopes (MAGIC) detected the gamma-ray afterglow of GRB 190114C, which can constrain microscopic parameters of the shock-heated plasma emitting non-thermal emission.
Focusing on the early afterglow of this event, we numerically simulate the spectrum and multi-wavelength light curves with constant and wind-like circumstellar medium using a time-dependent code.
Our results show that the electron acceleration timescale at the highest energies is likely shorter than 20 times the gyroperiod to reproduce the GeV gamma-ray flux and its spectral index
reported by {\it Fermi}. 
This gives an interesting constraint on the acceleration efficiency for Weibel-mediated shocks.
We also constrain the number fraction of non-thermal electrons $f_{\rm e}$, and the temperature of the thermal electrons. 
The early optical emission can be explained by the thermal synchrotron emission with $f_{\rm e} \lesssim 0.01$. On the other hand, the X-ray light curves restrict efficient energy transfer from protons to the thermal electrons, and $f_{\rm e}\sim1$ is required if the energy fraction of the thermal electrons is larger than $\sim10$\%.  
The parameter constraints obtained in this work give important clues to probing plasma physics with relativistic shocks.
\end{abstract}

\keywords{acceleration of particles --- gamma-rays: bursts
 --- radiation mechanisms: non-thermal}

\maketitle

\section{Introduction}
\label{sec:intro}
Gamma-ray burst (GRB) 190114C at redshift $z=0.4245$ is the first
gamma-ray burst detected with imaging atmospheric Cherenkov telescopes (IACTs).
\citet{mag19a} reported gamma-ray detection in the energy range of 0.2--1 TeV from 62 s to 2454 s after the trigger by the {\it Swift}-BAT. The spectral component in this energy range is naturally interpreted as the synchrotron self-Compton (SSC) emission from the afterglow caused by electrons accelerated at a blastwave \citep[][hereafter, MAGIC-MWL paper]{mag19},
because the photon energy is significantly larger than the maximum photon energy expected by synchrotron radiation
\citep[see also the case in GRB 130427A,][]{ack14}.
The SSC interpretation has been supported
by \citet{der19,fra19a,wan19,zha20}.

The SSC component in the early afterglow
uniquely provides the physical information
of the external shock in the early stage
(see \S \ref{sec:deg}).
In this paper, adopting the time-dependent code
in \citet{fuk17}, we simulate the broadband emission
of the afterglow in GRB 190114C.
We focus on the microscopic parameters
for the particle-acceleration in the relativistic shock,
especially the particle acceleration timescale,
the number fraction of the accelerated electrons,
and the temperature of the thermal electrons.
Although many studies on this topic have discussed
from a theoretical point of view \citep[e.g.,][]{spi08,lem11,sir11,sir13,kum15},
the particle-acceleration process and energy transfer from protons
to electrons in GRB afterglows
are not understood yet.
The multi-wavelength observations of GRB 190114C afterglow
can bring us hints for the acceleration mechanism.

In \S \ref{sec:method}, we explain our numerical method,
and review the parameter degeneracy in the afterglow modeling
in \S \ref{sec:deg}.
We show our results for spectrum and light curves in \S \ref{sec:results}.
In \S \ref{sec:TS}, we discuss thermal synchrotron emission
in the early afterglow, from which we can constrain
the heating efficiency of the thermal electrons
or the number fraction of accelerated electrons.
\S \ref{sec:sum} is devoted to summary.

\section{Numerical Methods}
\label{sec:method}
The afterglow emission is typically attributed to radiation from a shocked shell relativistically propagating in the circumstellar medium (external forward shock emission). Electrons accelerated at the shock (non-thermal electrons) emit synchrotron photons in a magnetic field amplified in the downstream. The non-thermal electrons up-scatter such synchrotron photons as well, and this process is called SSC emission. In this paper, the temporal evolution of the afterglow emission by the above two processes is calculated as follows.

\subsection{Method I: Time-Dependent Calculations}
We primarily adopt the numerical code in \citet{fuk17}
(hereafter F17).
The code follows the temporal evolutions of the bulk motion of the shocked-shell, magnetic field, and electron and photon energy distributions in the shell. The shell is assumed homogeneous within the shell width $\Delta R$. Physical processes addressed in the code are photon production and particle cooling by synchrotron and inverse-Compton with the Klein--Nishina effect, photon absorption by synchrotron self-absorption and $\gamma \gamma\rightarrow e^+e^-$ pair production, secondary pair injection, adiabatic cooling, and photon escape from the shell.
Integrating the escaped photons over the shell surface, we obtain the spectral evolution for an observer with effects of the Doppler beaming and the curvature of the emission surface to address the photon arrival time.

Given the density of the circumstellar medium $n$ and the
the bulk Lorentz factor of the shell $\Gamma=1/\sqrt{1-\beta_{\rm sh}^2}$
at a radius $R$,
we can follow the evolution of the shell mass with
\begin{equation}
\dot{M}=4 \pi R^2 c \beta_{\rm sh} n m_{\rm p}.
\end{equation}
The shock jump condition provides the energy injection into the shell.
The total energy in the shell frame $E'_{\rm sh}$ evolves with
the mass loading, radiative cooling, and adiabatic cooling.
The evolution of $\Gamma$ is calculated from the energy conservation
\begin{equation}
\Gamma E'_{\rm sh}=E_0+Mc^2-E_{\rm rad},
\label{eq:Gam}
\end{equation}
where $E_0$ is the total energy initially released,
and $E_{\rm rad}$ is the energy escaped from the shell as radiation.

In each time step, we add magnetic energy by assuming that the energy fraction $\epsilon_B$ of the downstream dissipated energy is converted into magnetic energy. The energy distribution of non-thermal electrons at injection is estimated by using the standard parameters: the energy fraction $\epsilon_{\rm e}$, and the number fraction $f_{\rm e}$ (see F17 for details).
The injection spectrum is assumed as a single power-law with an index $p$, minimum Lorentz factor $\gamma_{\rm m}$, and an exponential cutoff at $\gamma_{\rm max}$.  
The value of $\gamma_{\rm max}$ is obtained from the balance between the acceleration time and cooling time as $\eta \gamma_{\rm max} m_{\rm e} c/(e B)=t_{\rm c}$, where $\eta \geq 1$ is the acceleration efficiency parameter, and $t_{\rm c}$ is the cooling time due to synchrotron and inverse-Compton processes. 
Neglecting the inverse-Compton cooling with the simple approximation
$B^2/(8 \pi)=4 \Gamma^2 n m_{\rm p} c^2 \epsilon_B$,
the maximum Lorentz factor is approximated to be
\begin{equation}
\gamma_{\rm max} \approx \left( \frac{\pi}{2 \epsilon_B n m_{\rm p}} \right)^{1/4}
\left( \frac{3 e}{2 \eta \Gamma c \sigma_{\rm T}} \right)^{1/2}.
\label{gmax}
\end{equation}

The minimum Lorentz factor $\gamma_{\rm m}$ is numerically estimated,
taking into account $\gamma_{\rm max}$. 
In the limit of $\gamma_{\rm max} \to \infty$, we obtain the well-known formula
\begin{equation}
\gamma_{\rm m} \approx \frac{\epsilon_{\rm e}}{f_{\rm e}}
\frac{p-2}{p-1}(\Gamma-1)\frac{m_{\rm p}}{m_{\rm e}}.
\end{equation}

Given the electron injection spectrum $\dot{N}^{\rm inj}_{\rm e}(\gamma_{\rm e})$,
our numerical code follows the temporal evolutions of the energy distributions for electrons,
\begin{eqnarray}
\frac{\partial N_{\rm e}}{\partial t}&=&\frac{\partial}{\partial \gamma_{\rm e}}
\left[ \left( \dot{\gamma}_{\rm syn} + \dot{\gamma}_{\rm IC}
+ \dot{\gamma}_{\rm ad} - \dot{\gamma}_{\rm SSA} \right)
N_{\rm e}(\gamma_{\rm e}) \right] \nonumber \\
&&+\dot{N}^{\gamma \gamma}_{\rm e} + \dot{N}^{\rm inj}_{\rm e},
\label{diseq1}
\end{eqnarray}
and for photons,
\begin{eqnarray}
\frac{\partial N_{\rm \gamma}}{\partial t}=\dot{N}^{\rm syn}_{\gamma}
+ \dot{N}^{\rm IC}_{\gamma}- \dot{N}^{\gamma \gamma}_{\gamma}
- \dot{N}^{\rm SSA}_{\gamma}- \dot{N}^{\rm esc}_{\gamma},
\label{diseq2}
\end{eqnarray}
in the shell, where $\dot{\gamma}>0$ and $\dot{N}>0$
are the energy loss/gain rate normalized by the electron mass
and creation/annihilation rate, respectively,
for electrons (denoted with subscript e) and photons (subscript $\gamma$).
The superscripts, syn, IC, ad, SSA, $\gamma \gamma$, and esc,
express the contributions due to synchrotron emission,
inverse-Compton emission, adiabatic cooling, synchrotron self-absorption,
electron--positron pair creation, and photon escape, respectively.
From the density obtained from the jump condition
and total mass, we obtain the shell volume $V$
and the width as $\Delta R=V/(4 \pi R^2)$.
Then, the photon escape rate is calculated as
\begin{eqnarray}
\dot{N}^{\rm esc}_{\gamma}=\frac{c}{2W} N_{\rm \gamma}.
\end{eqnarray}
For electrons, we do not incorporate the escape effect,
because the mean free path is much shorter than the shell width
as will be discussed in \S \ref{sec:results}.
Alternatively, adiabatic cooling leads to a similar effect to the escape effect.
The details of other terms in Equations (\ref{diseq1}) and (\ref{diseq2})
are explained in \citet{fuk17}.

The electron energy distribution with the radiative cooling effect
can be approximated by a broken power-law \citep{mr97,sar98};
the electron spectrum has a low-energy cutoff at $\min(\gamma_{\rm m},\gamma_{\rm c})$
and break at $\max(\gamma_{\rm m},\gamma_{\rm c})$,
where $\gamma_{\rm c}$ corresponds to the cooling energy determined by
equality for the elapsed time and $t_{\rm c}$.
In the broken power-law approximation,
the spectral index above the break is $p+1$,
while the low-energy index is 2 for $\gamma_{\rm m}>\gamma_{\rm c}$
or $p$ for $\gamma_{\rm m}<\gamma_{\rm c}$.
As shown in F17,
our time-dependent numerical code yields
a smoothly curved electron spectrum.
The resultant photon spectrum also shows
a smoothly curved feature.
As a result, the different spectral shape
around the spectral peak leads to a different parameter set
from that with the conventional broken power-law approximation.

The flux obtained by this code can be different from
the conventional analytical approximation by a factor of two to three (F17).
The flux difference comes from the exact treatment in estimates of $\gamma_{\rm max}$
and $\gamma_{\rm m}$, the curved electron spectrum,
the flux estimate taking into account the equivalent arrival time surface,
and the time-dependent treatment with the effects of the adiabatic cooling
and inverse-Compton cooling.
As will be shown below, a larger $\epsilon_B$ compared to that
in MAGIC-MWL paper is required in our calculation.

\subsection{Method I\hspace{-.1em}I: Single-Zone Quasi-Steady Calculations}
In this work, we examine the results by an independent method, using another numerical code in \citet{mur11} (hereafter M11) with some modifications \citep[see also][for details]{zha+20}. 
In this method, we assume that the non-thermal electron distribution follows a power law. In the fast cooling regime ($\gamma_{\rm c}<\gamma_{\rm m}$), the steady-state electron distribution is used, which is given by
\begin{equation}
\frac{dN_{\rm e}}{d\gamma_{\rm e}}=\frac{t_{\rm c}}{\gamma_{\rm e}}\int_{\gamma_{\rm e}}d\gamma'_{\rm e} \frac{d\dot{N}_{\rm e}^{\rm inj}}{d\gamma'_{\rm e}},
\end{equation}
where $\dot{N}_{\rm e}^{\rm inj}$ is the injection rate of non-thermal electrons. In the slow-cooling case, we interpolate the injection spectrum and steady-state spectrum for $\gamma_{\rm m}<\gamma_e<\gamma_{\rm c}$. 
For dynamics, we use the Blandford--McKee solution with $R\approx 4\Gamma^2 ct_{\rm obs}/(1+z)$, and the results by this method agree with the analytical results \citep{sar98}. Compared with the results by F17, we find that the results are in agreement within a factor of two to three. 

One of the main differences comes from the fact that the single radiation zone is assumed without the integration over the equivalent time-arrival surface. There are other two notable differences. At low energies, heating due to the synchrotron self-absorption process can enhance the optical flux. In the GeV band, electromagnetics cascades can fill the dip between synchrotron and SSC components.

\section{Remarks on the Parameter Degeneracy}
\label{sec:deg}
Before showing our numerical results, we review the parameter degeneracy in the afterglow modeling.
The evolution of the bulk Lorentz factor $\Gamma$ is determined by the total energy $E_0$ and the density of the circumstellar medium;
the constant density $n_0$ or the wind profile
$n \propto A R^{-2}$.
However, as long as $E_0/n_0$ or $E_0/A$ is the same, different  values of $E_0$ lead to the same evolution of $\Gamma$ \citep{bla76}.

By adjusting the microscopic parameters, $f_{\rm e}$, $\epsilon_{\rm e}$, and $\epsilon_B$, we can obtain the same evolutions of the electron injection rate,
$\gamma_{\rm m}$, and the magnetic field $B$, for a different
value of $E_0$ \citep{eic05}.
Even if we obtain all the four spectral parameters,
$\varepsilon_{\rm a}$, $\varepsilon_{\rm m}$, $\varepsilon_{\rm c}$,
and $F_{\rm max}$ \citep[break energies due to synchrotron self-absorption,
$\gamma_{\rm m}$, and $\gamma_{\rm c}$, and the peak flux, respectively;][]{sar98}
at a certain observation time, we cannot determine all the five model parameters,
$E_0$, $n_0$ (or $A$), $\epsilon_{\rm e}$, $\epsilon_B$,
and $f_{\rm e}$\footnote{The injection index $p$ is directly obtained
from the photon spectral shape in ideal cases.}.

In other words, though we cannot determine the property of the ``thermal'' electrons,
whose number fraction is $1-f_{\rm e}$,
all the four practical parameters to yield the non-thermal emission,
the number of non-thermal electrons, $\gamma_{\rm m}$, $\gamma_{\rm c}$, and $B$,
can be uniquely determined by the observed synchrotron spectrum in an ideal case.
If we know all the four parameters above, inverse-Compton emission
can be automatically calculated without ambiguity in a single-zone model
\footnote{In a multi-zone model, the time differences between
the emission and scattering events
can play an important role in the light curve \citep[see e.g.,][]{mur11,che11}.}.
The inverse-Compton spectrum does not provide additional information
for the non-thermal electrons.
The parameter degeneracy is not solved by a detection
of the inverse-Compton component.

However, in the very early stage, all the four spectral parameters
for the synchrotron component, especially $\varepsilon_{\rm a}$, are rarely constrained.
Since the spectral parameters evolve monotonically in the standard afterglow model,
$\varepsilon_{\rm a}$ is usually extrapolated from radio observations
in the late stage \citep[see e.g.,][]{pan01}.
However, the spectral evolution can be affected by the jet break
in the late stage, and radio observations for some GRB samples
have shown inconsistent behavior with the standard afterglow model
\citep{kan19}.
The inconsistency may be resolved by the temporal evolutions of
the microscopic parameters, $\epsilon_{\rm e}$, $\epsilon_B$, and $f_{\rm e}$
\citep{iok06,mas14}.
Even for GRB 190114C, \citet{mis19} claimed that
a model with the evolving microscopic parameters agrees with
their long-term radio/mm observations.

If we assume constant values for the microscopic parameters, we should focus
on the spectrum in a limited time interval rather than
the entire spectral evolution.
Therefore, the inverse-Compton component provides a unique information
constraining the model parameters
in the very early stage of an afterglow not using the observational
data in the late stage.

Though the parameters degenerate with $f_{\rm e}$,
common values of $E_0 f_{\rm e}$,
$n_0 f_{\rm e}$ (or $A f_{\rm e}$), $\epsilon_{\rm e}/f_{\rm e}$,
and $\epsilon_B/f_{\rm e}$ yield
an identical model for different values of $f_{\rm e}$.
While the possible constraint on $f_{\rm e}$ will be
discussed in section \ref{sec:TS},
we fix $f_{\rm e}$ as 0.3 in our calculation with F17, which provides a reasonable value
for $E_0$.

We also have an additional microscopic parameter
$\eta$, which adjusts the maximum energy of accelerated electrons.
Particle-in-cell (PIC) simulations \citep[e.g.,][]{sir13}
show that the particle acceleration at a relativistic shock
is a diffusive process, in which $\gamma_{\rm max} \propto t^{1/2}$.
The simulation results imply
$\eta \approx r_{\rm L}/\lambda_{\rm min}$, where $r_{\rm L}$ is the particle's Larmor
radius, and $\lambda_{\rm min}$ is the minimum wavelength of plasma turbulence.
In this paper, we will constrain the value of $\eta$
with our numerical models, and discuss the consistency
with the PIC simulation result.

\section{Afterglow Spectrum and light curves}
\label{sec:results}
Focusing on the first $\sim 1000$ s,
we show two models of the GRB 190114C afterglow emission:
the constant circumstellar medium (ISM model)
and the wind-like circumstellar medium (wind model).
In Table \ref{table:para}, we summarize
the model parameters for the two models, respectively.

\begin{table*}[!htb]
  \caption{Model Parameters}
  \label{table:para}
  \centering
  \begin{tabular}{ccccccccc}
\hline
    Model & $E_0$ & $\Gamma_0$ & $n_0$ & $A$ &  $p$  & $\epsilon_{\rm e}$ & $\epsilon_B$ & $f_{\rm e}$  \\
          &  [erg]  &  & [$\mbox{cm}^{-3}$] & & & & & \\
    \hline \hline
    ISM	(method I) & $10^{54}$ & $600$ & 1.0 & --- & $2.3$ & $0.06$ & $9.0 \times 10^{-4}$ & 0.3 \\
    Wind (method I)& $10^{54}$ & $300$ & --- & $0.1$ & $2.35$ & $0.08$ & $1.2 \times 10^{-3}$ & 0.3 \\
    \hline
    ISM (method I\hspace{-.1em}I) & $4 \times 10^{53}$ & --- & $0.3$ & --- & $2.3$ & $0.1$ & $1.0 \times 10^{-3}$ & 1.0 \\
    \hline
  \end{tabular}
\end{table*}

As mentioned in the previous section, we adopt $f_{\rm e}=0.3$, which yields
$E_0 \simeq 10^{54} (f_{\rm e}/0.3)^{-1}$ erg in our modeling.
This value is reasonably larger than the prompt gamma-ray energy
$2.5 \times 10^{53}$ erg.

The optical flux at $t_{\rm obs}\sim 60$~s detected with
{\it Swift}/UVOT is too bright to be explained by
the forward shock emission.
So the optical emission at this stage may be dominated
by the reverse-shock component,
emission from a shock propagating inside
the ejecta coming from the central engine.
As this component contributes as seed photons
for inverse-Compton scattering,
we take into account the reverse-shock component
by manually adding a photon field in the shocked shell.
The spectrum of the reverse-shock component
is assumed to be the Band function \citep{ban93}
with the low-energy index $\alpha=-1$
and high-energy index $\beta=-2.5$.
The peak energy for the Band function is
adjusted as $\sim 10^{-2}$ eV in the shell rest frame
to reproduce the optical observation.

\subsection{ISM Model}
\begin{figure}[!t]
\centering
\epsscale{1.1}
\plotone{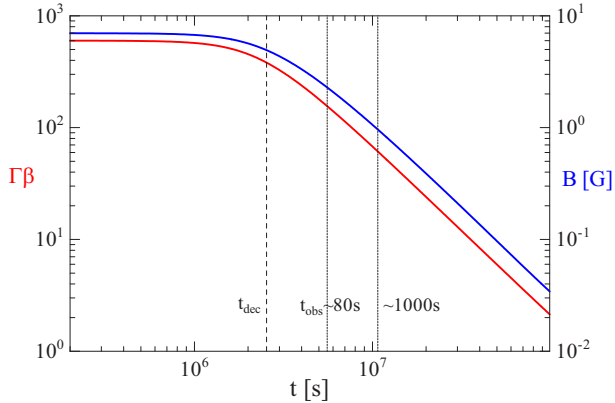}
\caption{Evolutions of the bulk Lorentz factor (red, left axis)
and the magnetic field (blue, right axis) for the ISM model.
The time $t$ is measured in the central engine rest frame.
The vertical dashed line corresponds to the deceleration time,
and the two vertical dotted lines indicate $\sim 80$ s
and $\sim 1000$ s as the observation times on Earth.
\label{fig:GbISM}}
\end{figure}

In Figure \ref{fig:GbISM}, we plot the evolutions of $\Gamma$ and $B$
versus time $t$ in the central engine rest frame for the ISM model.
The deceleration time, from which the shock starts to decelerate, is
\begin{eqnarray}
t_{\rm dec}&=&\frac{1}{c}
\left( \frac{3E_0}{4\pi n_0 m_{\rm p}c^2\Gamma_0^2} \right)^{1/3} \\
&\simeq& 2.5 \times 10^6 \left( \frac{E_0}{10^{54}~\mbox{erg}} \right)^{1/3}
\left( \frac{n_0}{1~\mbox{cm}^{-3}} \right)^{-1/3}
\left( \frac{\Gamma_0}{600} \right)^{-2/3} \mbox{s}. \nonumber \\
\end{eqnarray}
The time $t$ can be approximately transformed into the observer time as
$t_{\rm obs}=(1+z)t/(4 \Gamma^2)$ \citep{sar98},
though the emission detected by an observer at a certain time
is the superposition of photons emitted at different times
from different latitudes.
In Figure \ref{fig:GbISM}, we indicate the corresponding observer
times of $t_{\rm obs}=80$ s and 1000 s by dotted lines.
Most afterglow photons we discuss here are emitted
between the two dotted lines.

\begin{figure}[!t]
\centering
\epsscale{1.1}
\plotone{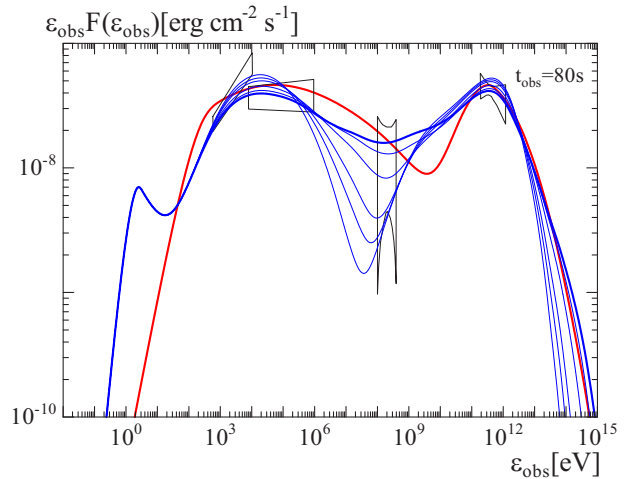}
\caption{Model spectra at $t_{\rm obs}=80$ s for the ISM model.
The observation data of {\it Swift} XRT and Fermi LAT are
taken from \citet{aje20},
and the MAGIC date, which are corrected for
attenuation caused by the extragalactic background light,
is taken from MAGIC-MWL paper.
We adopt the acceleration efficiency parameter $\eta=1$
for the thick solid line.
The blue lines are results with Method I,
and the red line is obtained with Method I\hspace{-.1em}I.
The blue thin lines show model spectra with $\eta=10$,
$100$, $1000$, $3000$, and $10,000$, decreasing the 100 MeV flux
(from top to bottom).
The spectral peak at $\sim 1$--10 eV is the ``reverse-shock'' component
we set manually.
\label{fig:SpISM}}
\end{figure}

\begin{figure}[!t]
\centering
\epsscale{1.1}
\plotone{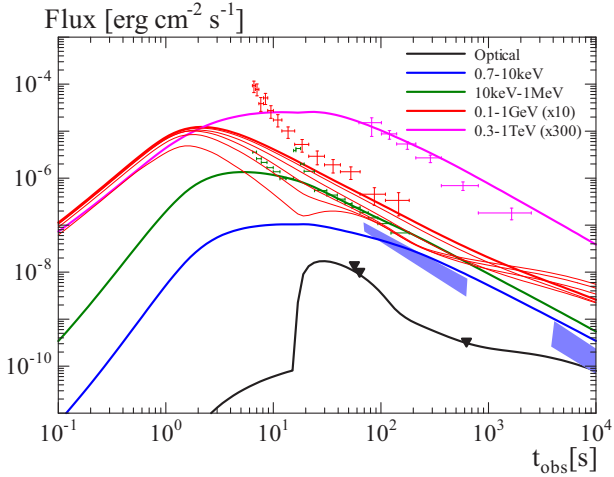}
\caption{Model light curves for the ISM model.
The observation data are
taken from \citet{aje20} (0.7--10 keV, 10 keV--1 MeV, 0.1--1 GeV)
and MAGIC-MWL paper (optical, 0.3--1 TeV).
The optical data are corrected for extinction
due to the host and our Galaxy.
The TeV gamma-ray data are also corrected for
attenuation caused by the extragalactic background light.
The thick lines show the model with $\eta=1$.
The red thin lines show the light curves for 0.1--1 GeV
with $\eta=10$, $100$, $1000$, $3000$, and $10,000$, decreasing the flux.
The optical bump at $\sim 20$--100 s is the ``reverse-shock'' component
we set manually.
\label{fig:LCISM}}
\end{figure}

The reverse-shock component is promptly added
at $t=4.9 \times 10^6$ s,
which roughly corresponds to $t_{\rm obs}=50$ s, with the energy density
of $0.7~\mbox{erg}~\mbox{cm}^{-3}$
and the spectral peak energy $2.6\times 10^{-2}$ eV in the shell rest frame.
Those parameters are chosen to reproduce the optical light curve
(see Figure \ref{fig:LCISM}).
While we cannot determine the spectral peak energy from the optical light curve alone,
our choice well restricts the effect of this additional component to the optical range
as shown in Figure \ref{fig:SpISM}.
The injected photons gradually escape from the shell with a timescale of
$t/(6 \Gamma)$ in the shell rest frame.
With the curvature effect on the dispersion of the photon arrival time,
the resultant reverse-shock light curve shows a smooth behavior
as shown in Figure \ref{fig:LCISM}.

Figure \ref{fig:SpISM} shows the afterglow spectrum
at $t_{\rm obs}=80$ s, when both MAGIC and {\it Fermi}
detected signals.
As mentioned above, the spectrum obtained with Method I
deviates from the conventional broken power-law formula
especially around the spectral peak.
The obtained flux is lower
than the analytical formula in F17 by a factor of 2.3 and 1.6
at 10 keV and MeV,
respectively. As the analytical formula neglects the inverse-Compton
emission, those flux differences are not so large.

The thick line is the case with $\eta=1$,
which realizes the theoretically highest value of $\gamma_{\rm max}$.
In this case, the synchrotron emission by
the electrons with a Lorentz factor close to $\gamma_{\rm max}$
is dominant in 0.1--1 GeV.
The seed photons for the inverse-Compton component
detected with MAGIC are dominated by the synchrotron photons
from the forward shock, rather than the reverse-shock component
at the optical energy range.

In Figure \ref{fig:SpISM}, we also plot the model spectrum
(red line) obtained by Method I\hspace{-.1em}I.
The model parameters are shown in Table \ref{table:para} as ISM (Method I\hspace{-.1em}I).
Although the curvature of the synchrotron spectrum around MeV
is different from the spectrum obtained by the time-dependent code,
we have obtained similar parameter sets as
$f_{\rm e} E_0 \sim 10^{53.5}$ erg, $f_{\rm e} n_0 \sim 0.3$,
$\epsilon_{\rm e}/f_{\rm e} \sim 0.1$,
and $\epsilon_B/f_{\rm e} \sim 10^{-3}$.
Thus, our time-dependent results by F17 are supported by an independent
calculation.

The multi-wavelength light curves
are well reproduced as shown in Figure \ref{fig:LCISM},
although the initial 0.1--1 GeV emission for $t_{\rm obs} \lesssim 10$ s,
to which the prompt component can contribute,
deviates from the model light curves.
At $t_{\rm obs} \sim 10^3$ s, the model fluxes of 0.3--1 TeV gamma-rays
and X-rays are slightly brighter than the observed data.
To reconcile those,
we may need the evolutions of microscopic parameters
in the framework of our model.
A parameter set of higher $\epsilon_B$ but lower $\epsilon_{\rm e}$
and $f_{\rm e}$ could realize the lower inverse-Compton flux
maintaining the synchrotron flux at $t_{\rm obs} \sim 10^3$ s.

As we decrease $\gamma_{\rm max}$ by increasing $\eta$
(see thin blue lines in Figure \ref{fig:SpISM}),
the 0.1--1 GeV emission is mostly coming from
inverse-Compton emission.
Since the error in the 0.1--1 GeV flux is large,
a very large $\eta$ seems acceptable from Figure \ref{fig:SpISM}.
However, the hard spectra for $\eta \gtrsim 1000$
do not agree with the photon index $\sim -2$
reported in \citet{aje20},
and the energy-integrated fluxes in Figure \ref{fig:LCISM}
are inconsistent with the cases of $\eta \gtrsim 100$.

The ion skin depth characterized by the proton plasma frequency
in the downstream is
\begin{eqnarray}
c/\omega_{\rm pp}&=&\sqrt{\frac{\Gamma m_{\rm p} c^2}{4 \pi e^2 (4 \Gamma n_0)}}
=\sqrt{\frac{m_{\rm p} c^2}{16 \pi e^2 n_0}} \\
&\simeq& 1.1\times 10^7 \left( \frac{n_0}{1~\mbox{cm}^{-3}} \right)^{-1/2}~\mbox{cm},
\end{eqnarray}
which is independent of the bulk Lorentz factor $\Gamma$.
The shortest wavelength mediated by the ion-Weibel instability
can be expressed as $\lambda_{\rm min}=\alpha c/\omega_{\rm pp}$,
where the dimensionless parameter $\alpha \sim 10$ \citep{ruy18}.
The parameter $\eta$ is inferred as $\sim r_{\rm L}/\lambda_{\rm min}$ by PIC simulations
\citep{sir13}.
This factor is estimated for electrons with the maximum 
Lorentz factor by using Equation (\ref{gmax}) and the expression of the acceleration time \citep[see also Equation (1) of][]{om19} as
\begin{eqnarray}
\eta_{\rm Weibel}&\approx&\frac{1}{\Gamma}
\frac{1}{(2 m_{\rm p})^{\frac{5}{6}} \epsilon_B^{\frac{1}{2}}}
\left( \frac{m_{\rm e}}{\alpha} \right)^{\frac{2}{3}}
\left( \frac{\pi}{n_0} \right)^{\frac{1}{6}}
\left( \frac{3 e}{c \sigma_{\rm T}} \right)^{\frac{1}{3}}\\
&\simeq& 100 \left( \frac{\Gamma}{156} \right)^{-1}
\left( \frac{\alpha}{7} \right)^{-\frac{2}{3}}
\left( \frac{\epsilon_B}{9 \times 10^{-4}} \right)^{-\frac{1}{2}} \nonumber \\
&&\times \left( \frac{n_0}{1~\mbox{cm}^{-3}} \right)^{-\frac{1}{6}},
\label{etamin}
\end{eqnarray}
where $\Gamma=156$ is the value
at $t=5.6 \times 10^6$ s ($t_{\rm obs}\simeq 80$ s)
in our simulation.

The value in Equation (\ref{etamin}) is based on the analytical approximation.
Given a value of $\eta$,
our simulation directly provides the magnetic field,
and $\gamma_{\rm max}$ obtained numerically:
$B=2.3$ G and
$\gamma_{\rm max}=8.7 \times 10^7$ ($7.8 \times 10^6$)
for $\eta=1$ (100) at $t=5.6 \times 10^6$ s.
The Larmor radius of the maximum-energy electrons
is $6.5 \times 10^{10}$ cm
($5.8 \times 10^{9}$ cm) for $\eta=1$ ($100$).
We can obtain a value of $\eta$ consistent
with $\eta \sim r_{\rm L}/\lambda_{\rm min}$ adjusting the parameter $\alpha$ as
\begin{eqnarray}
\eta\simeq 100 \left( \frac{\alpha}{5} \right)^{-1} \left( \frac{r_{\rm L}}{5.8\times 10^9 \mbox{cm}} \right)
\left( \frac{n_0}{1~\mbox{cm}^{-3}} \right)^{1/2}.
\end{eqnarray}
The fiducial value of $\alpha \sim 10$
can barely realize the required limit $\eta \sim 100$ at the maximum energy,
but to achieve the ideal value $\eta=1$,
a very large $\alpha \sim 5000$ is required.
Note that the Larmor radius for $\gamma_{\rm m}=1.7 \times 10^4$
is comparable to the ion skin depth,
namely shorter than $\lambda_{\rm min}$ for $\alpha \gtrsim 10$.
The acceleration processes around $\gamma_{\rm m}$
and $\gamma_{\rm max}$ may be different,
though we have assumed a single power-law injection.

Our parameter values of $\epsilon_{\rm e}/f_{\rm e}=0.2$
and $\epsilon_B/f_{\rm e}=3 \times 10^{-3}$
are significantly larger than those in the afterglow model
in MAGIC-MWL paper, $\epsilon_{\rm e}/f_{\rm e}=0.07$ and
$\epsilon_B/f_{\rm e}=8 \times 10^{-5}$.
Adopting the simple analytical formulae for the spectral break energies
neglecting the inverse-Compton cooling (see F17),
the parameter set of the model
in MAGIC-MWL paper provides $\varepsilon_{\rm c} \simeq 170~\mbox{keV}
\gg \varepsilon_{\rm m} \simeq 60$ eV 
at $t_{\rm obs}=80$ s.
Our numerical calculation with the same parameter set as that in MAGIC-MWL paper
leads to a dimmer synchrotron flux than
the observed one by a factor of $\sim 10$.
Even with the analytical formulae in F17, the MAGIC-MWL parameters
yield $\varepsilon_{\rm obs}F(\varepsilon_{\rm obs})
\simeq 10^{-8}~\mbox{erg}~\mbox{cm}^{-2}~\mbox{s}^{-1}$ at $\varepsilon_{\rm obs}=$MeV,
which is lower than the observed flux by a factor of $\sim 3$.
Thus, we need to adopt larger values of $\epsilon_{\rm e}/f_{\rm e}$
and $\epsilon_B/f_{\rm e}$ in our numerical method.
Those requirements are similar for the results with Method I\hspace{-.1em}I as well.
In both of the methods,
model fitting with $\epsilon_{B}/f_{\rm e} \ll 10^{-3}$
is difficult to reproduce both the synchrotron and SSC fluxes.

The analytical estimate for our parameter set in F17 gives us
$\varepsilon_{\rm c} \sim 3 \varepsilon_{\rm m} \sim 2$ keV.
The strong magnetic field leads to close values of $\varepsilon_{\rm c}$
and $\varepsilon_{\rm m}$.
Our numerical result
shows a smoothly curved spectrum so that it is hard to identify
the break energies of $\varepsilon_{\rm m}$ and $\varepsilon_{\rm c}$.
The synchrotron peak in Figure \ref{fig:SpISM} is slightly
higher than the analytical estimate of $\varepsilon_{\rm c}$
(see F17 for the detailed differences from the analytical formulae).
To keep a high flux above $\varepsilon_{\rm c}$, especially at GeV,
a small index of $p<2.5$ is required in our model.

\citet{aje20} concluded that the X-ray spectral break
($\sim 5$ keV) is due to $\varepsilon_{\rm c}$.
However, the analytical light curve for $\varepsilon_{\rm m}
<\varepsilon_{\rm obs}<\varepsilon_{\rm c}$ is
shallower ($\propto t^{3(1-p)/4}$) than the observed XRT light curve ($\propto t^{-1.3}$).
Our X-ray light curve also shows a slightly shallower decay
than the observed one.

\subsection{Wind Model}
\citet{aje20} claimed that the wind model can reconcile
the XRT light curve and
the spectral--temporal closure relation at $\varepsilon_{\rm m}
<\varepsilon_{\rm obs}<\varepsilon_{\rm c}$.
However, the wind model in MAGIC-MWL paper disagrees with
the high flux of the early MAGIC light curve.
\citet{fra19} proposed a possible transition from
wind-like medium to ISM-like medium at $t_{\rm obs} =300$--400 s.
Here, we also test the wind model with our numerical code.
The difference of flux from the analytical estimate
is similar to the ISM case.

\begin{figure}[!t]
\centering
\epsscale{1.1}
\plotone{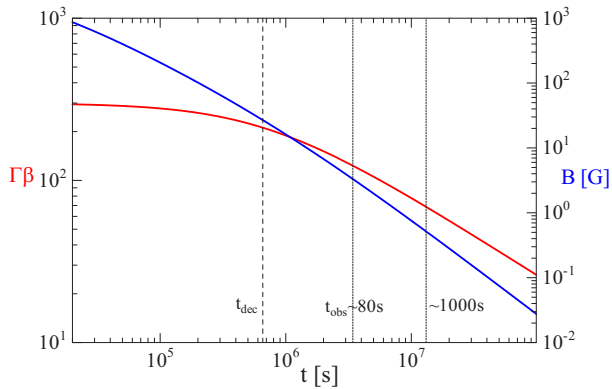}
\caption{Evolutions of the bulk Lorentz factor (red, left axis)
and the magnetic field (blue, right axis) for the wind model.
The time $t$ is measured in the central engine rest frame.
The vertical dashed line corresponds to the deceleration time,
and the two vertical dotted lines indicate $\sim 80$ s
and $\sim 1000$ s as the observation time on earth, respectively.
\label{fig:GbWind}}
\end{figure}

While a constant density was assumed in F17,
we can simulate the afterglow in a wind-like circumstellar environment
with the same numerical code adopting a density profile
\begin{eqnarray}
n=3.0 \times 10^{35} A R^{-2}~\mbox{cm}^{-3},
\end{eqnarray}
where the dimensionless parameter $A$
is interpreted by the mass-loss rate of the progenitor star as $10^{-5}A~M_\odot~\mbox{yr}^{-1}$
with the wind velocity of $10^3~\mbox{km}~\mbox{s}^{-1}$.
The evolutions of the bulk Lorentz factor $\Gamma$
and the magnetic field $B$ are shown in Figure \ref{fig:GbWind}.
In the asymptotic region ($t\gtrsim 10^7$ s),
the bulk Lorentz factor agrees with the analytical estimate of
$\Gamma \propto t^{-0.5}$,
while the magnetic field behaves as $B \propto t^{-1.4}$
slightly shallower than the analytical approximation $B \propto t^{-1.5}$.
At the time corresponding to $t_{\rm obs}\sim 80$ s,
$\Gamma$ deviates from the asymptotic power-law evolution,
which slightly affects the light curve behavior.

At $t=2.6\times 10^6$ s, which corresponds to $t_{\rm obs}\simeq 50$ s,
we add the reverse-shock component
with the energy density
of $1.7~\mbox{erg}~\mbox{cm}^{-3}$
and the peak energy of
$2.1 \times 10^{-2}$ eV in the shell rest frame.
The spectral indices are the same as the ISM case.

\begin{figure}[!t]
\centering
\epsscale{1.1}
\plotone{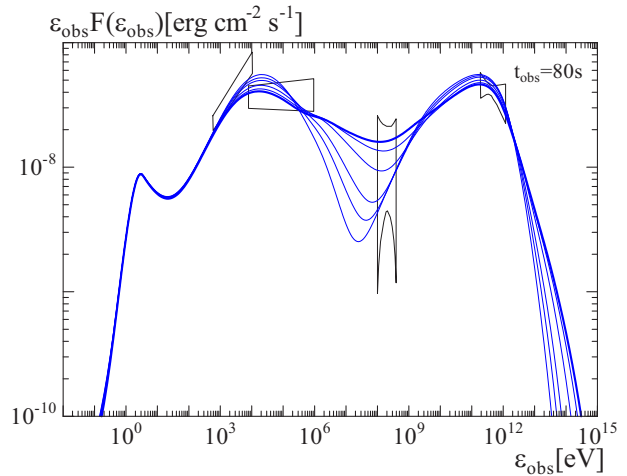}
\caption{Model spectra at $t_{\rm obs}=80$ s for the wind model.
We adopt the acceleration efficiency parameter $\eta=1$
for the thick solid line.
The thin lines show model spectra with $\eta=10$,
$100$, $1000$, $3000$, and $10,000$, decreasing the 100 MeV flux
(from top to bottom).
The spectral peak at $\sim 1$--10 eV is the ``reverse-shock'' component
we set manually.
\label{fig:SpWind}}
\end{figure}

\begin{figure}[!t]
\centering
\epsscale{1.1}
\plotone{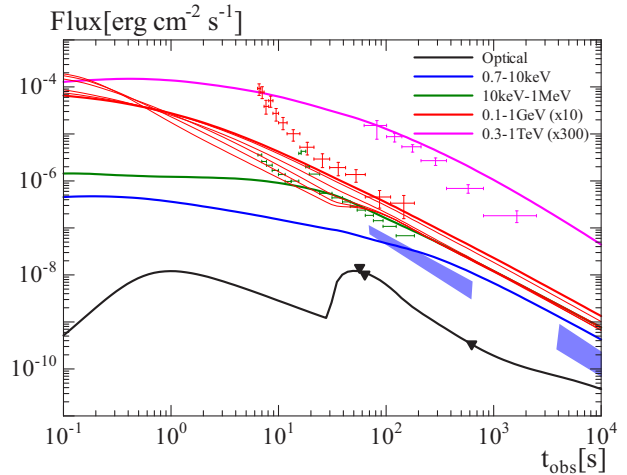}
\caption{Model light curves for the wind model.
The thick lines show the model with $\eta=1$.
The red thin lines show the light curves for 0.1--1 GeV
with $\eta=10$, $100$, $1000$, $3000$, and $10,000$, decreasing the flux.
The optical bump at $\sim 20$--100 s is the ``reverse-shock'' component
we set manually.
\label{fig:LCWind}}
\end{figure}

As shown in Figure \ref{fig:SpWind},
we obtain similar spectra to those for the ISM model
at $t_{\rm obs}=80$ s.
The constraint for $\eta$ for the wind model
is also similar to that for the ISM model;
$\eta\lesssim 100$ is required as shown in Figures \ref{fig:SpWind} and \ref{fig:LCWind}.
The combinations of the parameters  $\epsilon_{\rm e}/f_{\rm e}=0.27$
and $\epsilon_B/f_{\rm e}=4\times 10^{-3}$ in our models
are different from those for the wind model in MAGIC-MWL paper
($0.6$ and $10^{-4}$, respectively).
Similarly to the ISM model, the break energies of $\varepsilon_{\rm m}$
and $\varepsilon_{\rm c}$ reside in the X-ray band
at $t_{\rm obs}=80$ s.

Even with the wind model, we obtain similar light curves to those in the ISM model. Contrary to expectation suggested by \citet{aje20},
the X-ray model light curve is shallower than the XRT light curve
for $t_{\rm obs}<10^3$.
In our parameter set, the X-ray ($\sim 0.7$ keV) energy range is
still below $\varepsilon_{\rm m}$($<\varepsilon_{\rm c}$),
where the flux is supposed to be constant in the analytical
formula.
The early optical bump at $t_{\rm obs} \sim 1$ s in Figure \ref{fig:LCWind}
is originated from synchrotron emission from secondary electron--positron pairs
injected due to very high-density environment in the very early epoch.

In both the ISM and wind models,
the magnetic field and Lorentz factor at the time corresponding
to $t_{\rm obs} \simeq 80$ s
are similarly a few G and $\sim 100$, respectively
(see Figures \ref{fig:GbISM} and \ref{fig:GbWind}).
We can expect that those values do not largely depend on
models, though $\epsilon_B$ itself is not strongly constrained
because of the uncertainty in $f_{\rm e}$.

\section{Thermal Synchrotron Emission}
\label{sec:TS}
If a fraction of electrons are not injected into the acceleration process (i.e., $f_{\rm e}<1$), such ``thermal'' electrons also emit synchrotron photons.
\citet{eic05} pointed out that the thermal synchrotron
emission is expected in the radio wavelength at the early phase \citep[see also][]{war17,war18}, and it is naturally expected for non-relativistic or trans-relativistic shocks \cite{2020arXiv200502417S}.
However, the early-phase ($t\lesssim 10^2$ s) radio observation is challenging mission.
\citet{tom08} 
argued that the Faraday depolarization effect by the thermal electrons can be probed by the late-phase polarimetric observations at the radio frequencies above the synchrotron self-absorption frequency.
The relatively low polarization ($\sim 0.3$\%) of the GRB 171205A afterglow in the millimeter and submillimeter ranges may imply
$f_{\rm e}\sim 0.1$ \citep{ura19},
though it is difficult to set a robust lower-limit for $f_{\rm e}$.
While the radio observations have been focused in this issue,
\citet{res17} claimed that early X-ray and optical afterglow
could be dominated by the thermal synchrotron component.

Here we discuss an alternative interpretation with
the early thermal synchrotron signal for GRB 190114C.
If $f_{\rm e}$ is small enough, the early optical emission can be interpreted as the thermal synchrotron emission rather than the reverse-shock component as shown below.

Given the evolution of the bulk Lorentz factor as shown in Figure \ref{fig:GbISM} or \ref{fig:GbWind}, we can estimate the density of the thermal electrons $n_{\rm th}=(1-f_{\rm e})n_{\rm sh}$, where
\begin{eqnarray}
n_{\rm sh} \approx 4 \Gamma n,
\end{eqnarray}
and their temperature $T$ from the energy density,
\begin{eqnarray}
3T n_{\rm th}=\Gamma \left( 1+\epsilon_{\rm th} \frac{m_{\rm p}}{m_{\rm e}}
\right) n_{\rm th} m_{\rm e} c^2.
\end{eqnarray}
Here we have introduced a parameter $\epsilon_{\rm th}$, the fraction of the energy transferred from protons to thermal electrons.
Several PIC simulations of relativistic shocks \citep[e.g.,][]{spi08,sir11,kum15} have shown that the ion-Weibel instability significantly heats incoming electrons in the upstream.
The thermal electron energy density in the downstream reach the nearly equipartition with the ion energy density, so that we can expect $\epsilon_{\rm th}\sim 0.1$--$0.5$ according to those simulations.
A fraction of electrons are reflected at the shock entering the Fermi acceleration process.
The simulations by \citet{sir11} with $\Gamma=15$ are consistent with $f_{\rm e}\sim 0.02$ and $\epsilon_{\rm e} \sim 0.1$.

Given the uniform intensity $I_0(\varepsilon)$ at a stationary surface of radius $R$,
the luminosity is calculated as $L_0(\varepsilon)=4 \pi^2 R^2 I_0(\varepsilon)$.
If this surface is relativistically expanding with a Lorentz factor $\Gamma$,
the solid angle an observer can see
is $\Omega \approx \pi (R/\Gamma D_{\rm A})^2$ because of the relativistic beaming effect,
where $D_{\rm A}=D_{\rm L}/(1+z)$ is the angular diameter distance.
Using the luminosity distance $D_{\rm L}$ and the transformation
\begin{eqnarray}
I(\varepsilon) \approx \left( \frac{\Gamma}{1+z} \right)^3 I_0\left(
\varepsilon_0 \right),
\end{eqnarray}
within this solid angle, we obtain
the spectral flux for an observer as
\begin{eqnarray}
F_{\rm obs}(\varepsilon) \approx I(\varepsilon)\Omega
\approx \frac{\Gamma (1+z)}{4 \pi D^2_{\rm L}}
L_0\left( \varepsilon_0 \right),
\end{eqnarray}
where $\varepsilon_0=(1+z)\varepsilon/\Gamma$.

With the effect of synchrotron self-absorption,
the intensity of the thermal synchrotron emission is written as
\begin{eqnarray}
I_{\rm th}(\varepsilon)=\frac{2 \varepsilon^2 T}{c^2 h^3}
\left( 1-e^{-\tau(\varepsilon)} \right),
\end{eqnarray}
where $\tau(\varepsilon)=\alpha(\varepsilon) \Delta R$ is the optical depth.
In a thermal plasma, the absorption coefficient is
\begin{eqnarray}
\alpha(\varepsilon)=\frac{j_{\rm th}(\varepsilon) c^2 h^3}{2 \varepsilon^2 T},
\end{eqnarray}
where $j_{\rm th}(\varepsilon)$ is the thermal synchrotron emissivity calculated from the magnetic field and the electron energy distribution
\begin{eqnarray}
\frac{dn_{\rm th}}{d \varepsilon_{\rm e}}
=\frac{n_{\rm th}}{2 T^3} \varepsilon_{\rm e}^2
\exp{\left(-\varepsilon_{\rm e}/T\right)}.
\end{eqnarray}
The single-zone approximation with particle number conservation provides us the shell width in the shell rest frame as $\Delta R=R/(12\Gamma)$ and $R/(4\Gamma)$ for the ISM and wind cases, respectively.
Finally we obtain the thermal synchrotron flux for an observer as
\begin{eqnarray}
F_{\rm th}(\varepsilon) \approx \pi \frac{(1+z)^3}{\Gamma} \left( \frac{R}{D_{\rm L}} \right)^2
\frac{2 \varepsilon^2 T}{c^2 h^3}
\left( 1-e^{-\tau(\varepsilon_0)} \right).
\label{thflux}
\end{eqnarray}

\begin{figure}[!h]
\centering
\epsscale{1.1}
\plotone{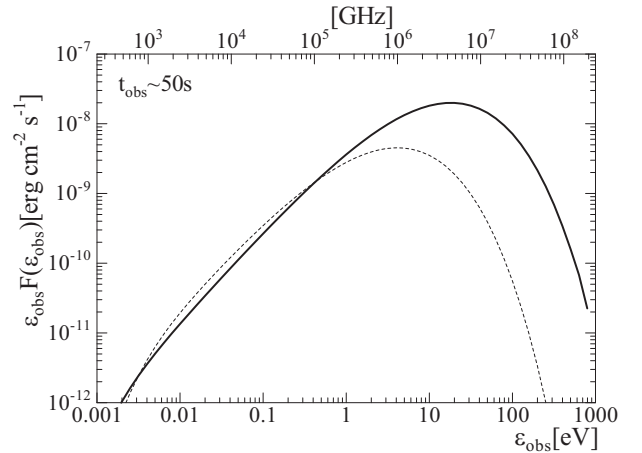}
\caption{Model spectra for the thermal synchrotron emission
at $t_{\rm obs}\sim 50$~s in the ISM model.
The evolutions of the Lorentz factor $\Gamma$ and magnetic field $B$
are the same as those in Figure \ref{fig:GbISM}.
We fix $f_{\rm e}=0.01$, and adopt $\epsilon_{\rm th}=0$ (solid)
and $6 \times 10^{-4}$ (dashed), respectively.
\label{fig:ThISMSP}}
\end{figure}

\begin{figure}[!h]
\centering
\epsscale{1.1}
\plotone{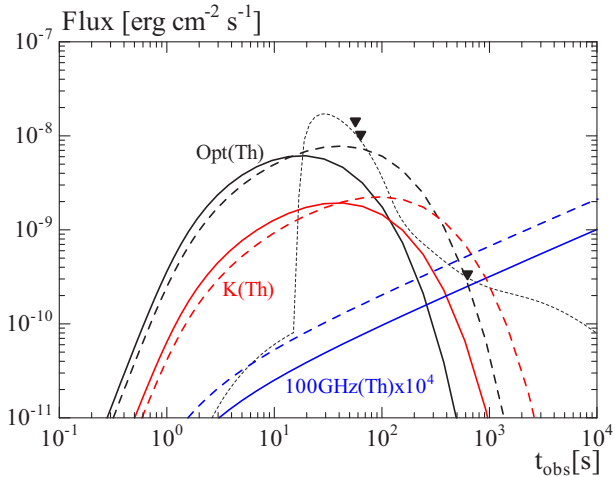}
\caption{Model light curves for the thermal synchrotron emission
in the ISM model.
The evolutions of the Lorentz factor $\Gamma$ and magnetic field $B$
are the same as those in Figure \ref{fig:GbISM}.
The black thin dashed line is the same non-thermal model
for the optical band in Figure \ref{fig:LCISM}.
The thermal optical (black), infrared (K band, red),
and radio (100 GHz, blue) are plotted.
We fix $f_{\rm e}=0.01$, and adopt $\epsilon_{\rm th}=0$ (solid)
and $6 \times 10^{-4}$ (dashed), respectively.
\label{fig:ThISMLC}}
\end{figure}

Given the Lorentz factor and magnetic field (see Figure \ref{fig:GbISM}),
Equation (\ref{thflux}) provides us the thermal synchrotron spectrum at an arbitrary time $t_{\rm obs}=(1+z)t/(4 \Gamma^2)$ as shown in Figure \ref{fig:ThISMSP}.
To make the thermal synchrotron flux comparable to the observed optical flux, we need $f_{\rm e} \lesssim 0.01$, namely the number density of the thermal electrons is required to be more than 100 times the non-thermal electron density.
The synchrotron self-absorption frequency in the examples in Figure \ref{fig:ThISMSP} is $\sim$ THz,
so that the polarization in radio band typically below 100 GHz should be greatly suppressed in those parameter sets.
In Figure \ref{fig:ThISMLC}, we plot the thermal synchrotron light curves with $f_{\rm e}=0.01$.
Even with a small value of $\epsilon_{\rm th}<10^{-3}$,
the thermal emission can reproduce the optical flux
at $t_{\rm obs}=50$--60~s without the reverse-shock component.

The characteristic point in the thermal model is the light curve crossing for the optical and infrared bands at $t_{\rm obs}\sim 100$~s (300~s) for $\epsilon_{\rm th}=0$ ($6 \times 10^{-4}$).
Such behavior is hard to be realized by non-thermal emission mechanisms.

\begin{figure}[!h]
\centering
\epsscale{1.1}
\plotone{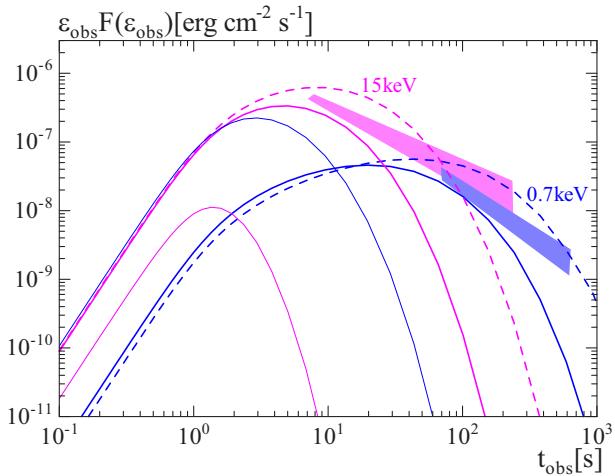}
\caption{X-ray Model light curves for the thermal synchrotron emission
in the ISM model.
The non-thermal X-ray fluxes ($\varepsilon_{\rm obs} F(\varepsilon_{\rm obs})$,
shaded regions) at 0.7 keV (blue)
and 15 keV (magenta) are estimated from the data in \citet{aje20}
assuming $\varepsilon_{\rm obs} F(\varepsilon_{\rm obs}) \propto \varepsilon_{\rm obs}^0$.
The evolutions of the Lorentz factor $\Gamma$ and magnetic field $B$
are the same as those in Figure \ref{fig:GbISM}.
The thermal synchrotron emission at 0.7 keV (blue) and 15 keV (magenta) are plotted
with parameter values of $f_{\rm e}=0.01$ \& $\epsilon_{\rm th}=6 \times 10^{-4}$ (thin solid),
$f_{\rm e}=0.3$ \& $\epsilon_{\rm th}=0.01$ (thick solid),
and $f_{\rm e}=0.3$ \& $\epsilon_{\rm th}=0.02$ (thick dashed).
\label{fig:ThXLC}}
\end{figure}

The value $f_{\rm e}=0.01$ requires a very large energy $E_0=3 \times 10^{55}$ erg, which can be regarded as a caveat of the interpretation by synchrotron emission from thermal electrons.
Even assuming a reasonable value of $f_{\rm e}=0.3$, we can constrain the parameter value of $\epsilon_{\rm th}$ from the early X-ray light curves.

\begin{figure}[!h]
\centering
\epsscale{1.1}
\plotone{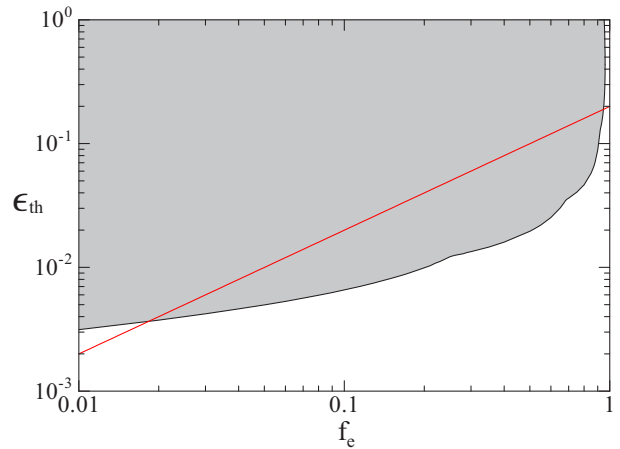}
\caption{The upper limits of $\epsilon_{\rm th}$ in the ISM model.
The parameter sets in the gray shaded region yield a brighter X-ray flux than observed. The red line shows the value of 
$\epsilon_{\rm e} = 0.06\;(f_{\rm e}/0.3)$ in our model.
\label{fig:ThUL}}
\end{figure}

Changing the value of $f_{\rm e}$, we obtain the upper limit of $\epsilon_{\rm th}$ from the constraints by the 0.7 keV and 15 keV light curves. The result is shown in Figure \ref{fig:ThUL}.
For a finite value of $1-f_{\rm e}$, a value of $\epsilon_{\rm th}$ larger than $10^{-2}$ is unlikely.
Only the case of $f_{\rm e} \simeq 1$ is acceptable
for $\epsilon_{\rm th}\gtrsim {10}^{-2}$.
The small $\epsilon_{\rm th}$ does not agree with the present PIC simulations of relativistic shocks.
Our results imply that the temperature of the thermal electrons should be not much larger than $\Gamma m_{\rm e} c^2$ if a dominant fraction of electrons remains thermal.
Some additional effects to make the thermal electrons reenter
the Fermi acceleration process are required, which increases the non-thermal fraction or makes all electrons be non-thermal population (i.e., $f_{\rm e} \simeq 1$).

In the wind model, the stronger magnetic field in the early stage
gives a more stringent upper limit for $\epsilon_{\rm th}$
than that in Figure \ref{fig:ThUL}.

\section{Summary and Discussion}
\label{sec:sum}
The inverse-Compton component detected with MAGIC telescopes from GRB 190114C uniquely constrains the magnetic field and non-thermal electron population at the early phase of the afterglow.
In this paper, using our time-dependent code (Method I) provided by F17, we reproduced the broadband spectrum and light curves of the early afterglow of GRB 190114C.
The flux ratio of the inverse-Compton component to the synchrotron component at $t_{\rm obs}\simeq 80$~s is consistent with the models with microscopic parameters of $\epsilon_{\rm e}/f_{\rm e} \sim 0.1$ and $\epsilon_B/f_{\rm e} \sim 10^{-3}$, irrespective of the models of the circumstellar environment (ISM or wind).
The independent numerical code (Method I\hspace{-.1em}I) also provides a result consistent with $\epsilon_B/f_{\rm e} \sim 10^{-3}$. 
The required magnetic field and Lorentz factor are a few gausses and $\sim 100$, respectively, at the time corresponding to $t_{\rm obs} \simeq 80$~s.

However, the observed decay index of the X-ray afterglow is slightly steeper than the model light curves.
The spectra shown in \citet{aje20} do not show a significant evolution of the break energy ($\sim 5$~keV) from $t_{\rm obs}= 68$~s to 627~s, which does not agree with both the standard ISM and wind models for both the fast ($\varepsilon_{\rm c}<\varepsilon_{\rm m}$) and slow ($\varepsilon_{\rm m}<\varepsilon_{\rm c}$) cooling cases.
Although this unexpected behavior of the break energy may require the time evolution of the microscopic parameters even for this time interval, the strength of the magnetic field can be expected not far different from our estimate at least at $t_{\rm obs} \simeq 80$~s (see \S \ref{sec:deg}).

The flux detected with {\it Fermi} above 0.1~GeV constrains the acceleration efficiency parameter as $\eta\lesssim100$, which adjusts the maximum electron Lorentz factor $\gamma_{\rm max}$.
The acceleration timescale is required shorter than $100/(2\pi)\sim 20$ times the gyroperiod.
The simple estimate $\eta \sim \gamma_{\rm max} m_{\rm e} c \omega_{\rm pp}/(10 e B)$ supported by the state-of-art PIC simulations seems to be marginally consistent with $\eta\sim100$.
If the actual $\eta$ is much smaller, the maximum energy of non-thermal electrons would need to be regulated by another mechanism rather than the diffusive process seen in the early stage of the Fermi acceleration in the PIC simulations.
For example, a large-scale MHD turbulence may play an important role in acceleration of the highest-energy electrons \citep{zha09,ino11,dem18,ter19}.
Note that the Larmor radius of electrons of $\gamma_{\rm m}$ may be shorter than the coherence length scale $\lambda_{\rm min}$ of turbulence required to make $\eta$ small enough.
This implies a different acceleration process for such low-energy electrons.
The entire particle-acceleration mechanism at relativistic shocks may be a compound one.

It is natural that only a fraction of electrons are accelerated by the shock and the rest of electrons remain as the thermal component, which is especially the case for mildly relativistic shocks.
The early optical and X-ray afterglow emissions constrain the non-thermal fraction $f_{\rm e}$ and the heating efficiency $\epsilon_{\rm th}$ of the thermal electrons.
Intriguingly, we found that the thermal synchrotron model with $f_{\rm e}=0.01$ and $\epsilon_{\rm th}=6 \times 10^{-4}$ can explain the early optical emission instead of the reverse-shock emission, although the required total energy becomes as large as $E_0=3 \times 10^{55}$ erg.
This model predicts a characteristic behavior of light curves -- light curve crossing for the optical and infrared bands.
However, this interpretation would contradict with the results of the PIC simulations for ultra-relativistic shocks.
The PIC simulations have shown fairly large values of $\epsilon_{\rm th}$, i.e., a significant fraction of electrons are heated.
In this case, if $f_{\rm e} \ll 1$, the thermal X-ray emission should contribute to the early afterglow. The absence of such a component in the X-ray light curves rules out values $\epsilon_{\rm th}\gtrsim 10^{-2}$ or indicates $f_{\rm e}\simeq 1$.
Those results provide us important clue to probing plasma physics with relativistic shocks.

\begin{acknowledgments}
We acknowledge useful suggestions by the anonymous referee.
This work is supported by the joint research program
of the Institute for Cosmic Ray Research (ICRR),
the University of Tokyo.
The work of K.M. is supported by the Fermi GI program 111180,
and NSF grant No.~AST-1908689. This work is also supported by JSPS KAKENHI Grant No.~20H01901 (K.M.) and No.~18H01245 (K.T.). 
\end{acknowledgments}


\begin{thebibliography}{}

\bibitem[Ackermann et al.(2014)]{ack14} 
Ackermann, M., Ajello, M., Asano, K., et al. 2014, Science, 343, 42
\bibitem[Ajello et al.(2020)]{aje20} 
Ajello, M., Arimoto, M., Axelsson, M., et al. 2020, \apj, 890, 9
\bibitem[Band et al.(1993)]{ban93}
Band, D., Matteson, J., Ford, L., et al. 1993, \apj, 413, 281
\bibitem[Blandford \& Mckee(1976)]{bla76}
Blandford, R. D., \& Mckee, C. F. 1976, PhFl, 19, 1130
\bibitem[Chen et al.(2011)]{che11}
Chen, X., Fossati, G., Liang, E. P., \& B\"ottcher, M. 2011, \mnras, 416, 2368
\bibitem[Demidem et al.(2018)]{dem18}
Demidem, C., Lemoine, M., \& Casse, F. 2018, \mnras, 475, 2713
\bibitem[Derishev \& Piran(2019)]{der19}
Derishev, E., \& Piran, T. 2019, \apjl, 880, L27
\bibitem[Eichler \& Waxman(2005)]{eic05}
Eichler, D., \& Waxman, E. 2005, \apj, 627, 861
\bibitem[Fraija et al.(2019)]{fra19a}
Fraija, N., Barniol Duran, R., Dichiara, S., \& Beniamini, P. 2019, \apj, 883, 162
\bibitem[Fraija et al.(2019)]{fra19}
Fraija, N., Dichiara, S., Pedreira, A. C., et al. 2019, \apjl, 879, L26
\bibitem[Fukushima et al.(2017)]{fuk17}
Fukushima, T., To, S., Asano, K., \& Fujita, Y. 2017, \apj, 844, 92 (F17)
\bibitem[Inoue et al.(2011)]{ino11}
Inoue, T., Asano, K., \& Ioka, K. 2011, \apj, 734, 77
\bibitem[Ioka et al.(2006)]{iok06}
Ioka, K., Toma, K., Yamazaki, R., \& Nakamura, T. 2009, \aap, 458, 7
\bibitem[Kangas \& Fruchter(2001)]{kan19}
Kangas, T., \& Fruchter, A. S. 2019, arXiv:1911.01938
\bibitem[Kumar et al.(2015)]{kum15}
Kumar, R., Eichler, D., \& Gedalin, M. 2015, \apj, 806, 165
\bibitem[Lemoine \& Pelletier(2011)]{lem11}
Lemoine, M., \& Pelletier, G. 2011, \mnras, 417, 1148
\bibitem[MAGIC Collaboration (2019)]{mag19a}
MAGIC Collaboration 2019, \nat, 575, 455
\bibitem[MAGIC Collaboration, et al.(2019)]{mag19}
MAGIC Collaboration, et al. 2019, \nat, 575, 459 (MAGIC-MWL)
\bibitem[Maselli et al.(2014)]{mas14}
Maselli, A., Melandri, A., Nava, L., et al. 2014, Science, 343, 48
\bibitem[M\'es\'aros \& Rees (1997)]{mr97}
M\'esz\'aros, P. \& Rees, M. J. 1997, ApJ, 476, 232
\bibitem[Misra et al.(2019)]{mis19}
Misra, K., Resmi, L., Kann, D. A., et al. 2019, arXiv:1911.09719
\bibitem[Murase et al.(2011)]{mur11}
Murase, K., Toma, K., Yamazaki, R.,
\& M\'esz\'aros, P. 2011, \apj, 732, 77 (M11)
\bibitem[Ohira \& Murase(2019)]{om19}
Ohira, Y., \& Murase, K. 2019, \prd, 100, 061301(R)
\bibitem[Panaitescu \& Kumar(2001)]{pan01}
Panaitescu, A., \& Kumar, P. 2001, \apj, 554, 667
\bibitem[Ressler \& Laskar(2017)]{res17}
Ressler, S. M., \& Laskar, T. 2017, \apj, 845, 150
\bibitem[Ruyer \& Fiuza(2018)]{ruy18}
Ruyer, C., \& Fiuza, F. 2018, \prl, 120, 245002
\bibitem[Samuelsson et al.(2020)]{2020arXiv200502417S} 
Samuelsson F., B{\'e}gu{\'e} D., Ryde F., Pe'er A., Murase K., 2020, \apj, 902, 148
\bibitem[Sari et al.(1998)]{sar98}
Sari, R., Piran, T., \& Narayan, R., 1998, \apjl, 497, L17
\bibitem[Sironi \& Spitkovsky (2011)]{sir11}
Sironi, L., \& Spitkovsky, A. 2011, \apj, 726, 75
\bibitem[Sironi et al. (2013)]{sir13}
Sironi, L., Spitkovsky, A., \& Arons, J. 2013, \apj, 771, 54
\bibitem[Spitkovsky (2008)]{spi08}
Spitkovsky, A. 2008, \apjl, 673, L39
\bibitem[Teraki \& Asano (2019)]{ter19}
Teraki, Y., \& Asano, K. 2019, \apj, 877, 71
\bibitem[Toma et al. (2008)]{tom08}
Toma, K., Ioka, K., \& Nakamura, T. 2008, \apjl, 673, L123
\bibitem[Urata et al. (2019)]{ura19}
Urata, Y., Toma, K., Huang, K., Asada K., Nagai, H.,
Takahashi, S., Petitpas, G., Tashiro, M., \& Yamaoka, K. 2019, \apjl, 884, L58
\bibitem[Wang et al. (2019)]{wan19}
Wang, X.-Y., Liu, R.-Y., Zhang, H.-M., Xi, S.-Q., \& Zhang, B. 2019, \apj, 884, 117
\bibitem[Warren et al. (2017)]{war17}
Warren, D. C., Ellison, D. C., Barkov, M. V., \& Nagataki, S. 2017, \apj, 835, 248
\bibitem[Warren et al. (2018)]{war18}
Warren, D. C., Barkov, M. V., Ito, H., Nagataki, S., \& Laskar, T. 2018, \mnras, 480, 4060
\bibitem[Zhang et al. (2020)]{zha20}
Zhang, H., Christie, I. M., Petropoulou, M., Rueda-Becerril, J. M., \& Giannios, D. 2020, \mnras, 496, 974
\bibitem[Zhang et al.(2020b)]{zha+20}
Zhang, B., Murase, K., Yuan, C., Kimura, S. S., \& M\'esz\'aros, P. 2020b, in preparation
\bibitem[Zhang et al.(2009)]{zha09}
Zhang, W., MacFadyen, A., \& Wang, P. 2009, \apj, 692, 240
\end{thebibliography}
\end{document}